\documentstyle[epsf]{mn2e}

\title[8 years of RE J1034+396]
{Searching for the trigger of the AGN QPO:\\ 8 years of RE J1034+396}

\author[M. Middleton, P. Uttley \& C. Done ]
{Matthew Middleton$^1$, Phil Uttley$^2$ \& Chris Done$^1$\\
$^1$Department of Physics, University of Durham, South Road, Durham
DH1 3LE,
UK\\
$^2$School of Physics and Astronomy, University of Southampton, Highfield, Southampton, SO17 1BJ, UK 
}

\pagerange{\pageref{firstpage}--\pageref{lastpage}} \pubyear{2009}

\begin{document}

\topmargin = -0.5cm

\maketitle

\label{firstpage}

\begin{abstract}

RE J1034+396 is one of the most extreme Narrow-line Seyfert 1s
detected thus far, showing the only quasi-periodic oscillation (QPO)
reliably detected in an Active Galactic Nucleus (AGN). Comparison with
similar spectral and timing properties observed in the black hole
X-ray binary (BHB) GRS~1915+105 suggests that RE J1034+396 is a
super-Eddington accretor. A more complete understanding of the
behaviour of RE J1034+396 can therefore lead to a unification of the
accretion physics between such extreme AGN and super-Eddington
BHBs. Here we report on our latest {\it XMM-Newton} observations of RE
J1034+396, which no longer show the QPO, indicating that this source
shows a non-stationary power-spectrum. We
use spectral and temporal analysis across all five \textit{XMM-Newton}
observations of the source to probe the evolution of the object. 
The combination of the shape of the fractional variability with energy and
the inferred velocity of absorbing material in the line-of-sight rules
out an absorption-only method of creating the QPO. Instead the 
periodically changing absorption may be produced by the QPO causing a
change in ionization state. 

We extend our analysis by including the covariance spectra which give
much better signal to noise than an rms spectrum.  These reveal a new
aspect of the QPO, which is that there is also a small contribution
from a soft component which is hotter than the soft excess seen in the
mean spectrum. Folding the lightcurve on the QPO period shows that
this component lags behind the hard X-rays. If this is due to 
re-processing then the lag corresponding to a light travel time
across $\sim$30R$_{g}$. Some of the remaining observations have
similar energy spectra and covariance spectra, but none of them show a
significant QPO, so we conclude that none of these features are the
trigger for the appearance of the QPO in this object.

\end{abstract}
\begin{keywords}  accretion, accretion discs -- X-rays: binaries, black hole
\end{keywords}

\section{Introduction}

Black holes should be entirely described by their mass and spin, but
their observational appearance is also determined by their mass
accretion rate.  Current models of the accretion flow, both thin disc
and hot flow solutions, show that the underlying physics should be
fairly scale invariant i.e. the properties of the flow should simply
scale mostly with mass for a given mass accretion rate in Eddington
units, $L/L_{Edd}$. Thus data from the well studied, stellar mass
black hole binary (BHB) sources can be simply scaled up to describe
the supermassive black holes in the center of AGN.  There is
considerable evidence to support this conjecture. The different
spectral states seen as a function of mass accretion rate in BHB may
well be seen in AGN (Pounds, Done \& Osbourne 1995; Vasudevan \&
Fabian 2007; Middleton, Done \& Gierli{\'n}ski 2007), and `fundamental-plane' type
relations between radio and X-ray luminosity hold across the
mass-scale (Merloni, Heinz \& di Matteo 2003; Falcke, K{\"o}rding \&
Markoff 2004; K{\"o}erding et al. 2007). The X-ray timing properties also support a simple
scaling argument, with both BHBs and AGN having a similar shape in
their broad-band power density spectrum (PDS), but shifted in
frequency due to the expected mass-scaling of the characteristic
variability time-scales (M$^{\rm c}$Hardy et al. 2004; Done \&
Gierli{\'n}ski 2005; M$^{\rm c}$Hardy et al. 2006).

\begin{figure*}
\begin{center}
\begin{tabular}{l}
 \epsfxsize=10cm \epsfbox{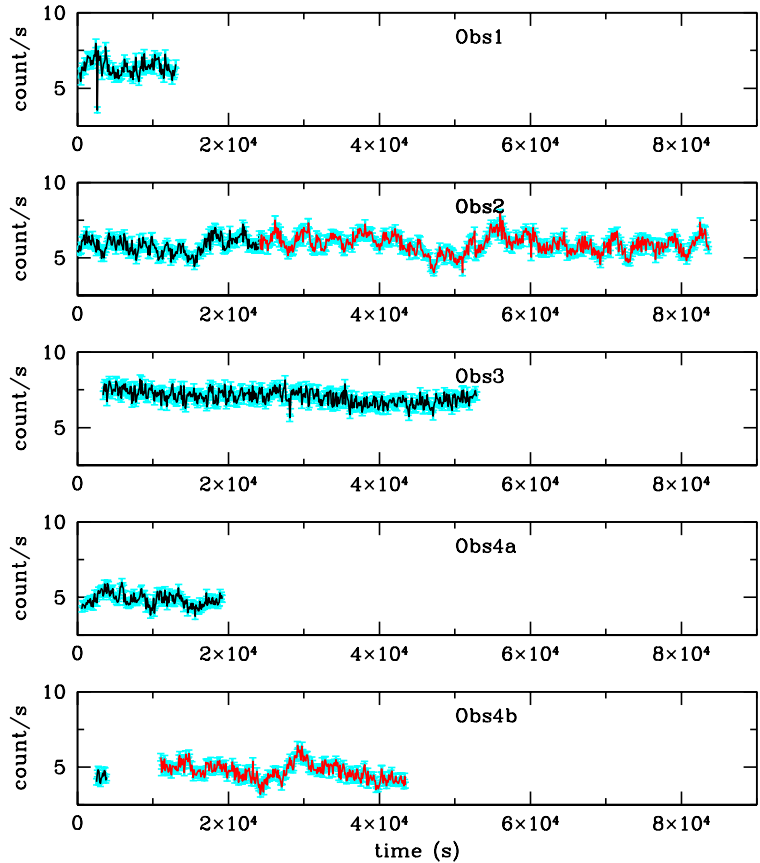}
\end{tabular}
\end{center}
\caption{Background subtracted, co-added lightcurves of all five observations of RE J1034+396 taken
with \textit{XMM-Newton}, binned on 100 s. Obs1: the first observation
(OBSID: 0109070101) of only 16~ks total duration and containing a large
amount of background flaring; Obs2: the important second observation
containing the QPO (OBSID: 0506440101 showing the `in-phase' segment
of $\sim$60~ks in red and the remaining lightcurve in black); Obs3: the DDT observation (OBSID: 0561580201) containing
$\sim$50~ks of good data ; Obs4a and 4b the follow-up observation taken
in AO9 and broken into two segments the first of which was heavily
dominated by background flaring (leaving only $\sim$ 20 ks of good
time) and the second containing moderate flaring episodes leaving ~30ks of continuous time (red).}
\label{fig:l}
\end{figure*}

However, BHBs also often show prominent low frequency quasi-periodic
oscillations (LFQPOs) in their X-ray lightcurves (e.g. Remillard \&
McClintock 2006). These have frequencies from 0.1-10~Hz, so for
typical AGN masses these are not detectable given the lightcurves
currently available (Vaughan \& Uttley 2005). Thus the single
significant AGN QPO, seen in the Narrow Line Seyfert 1 galaxy
RE~J1034+396 (Gierli{\'n}ski et al. 2008; confirmed using even more
robust techniques by Vaughan 2010) is most likely associated
instead with the higher frequency QPOs seen in BHBs. These are
associated only with very specific spectral states (very
high/intermediate), so are much less frequently observed
(e.g. Remillard et al. 2002; Remillard \& McClintock 2006). While
there is currently no reliable mass estimate for this AGN (though
recent work is converging on $\sim$1-4$\times$10$^{6}$M$_{\odot}$ : e.g. Bian
\& Huang 2010; Jin et al. in press), the best match in terms of mass scaled
frequency (and L/L$_{Edd}$) appears to be the 67~Hz QPO seen in the
probably super-Eddington BHB GRS 1915+105 (e.g. Middleton \& Done 2010; Morgan, Remillard \& Greiner 1997; Ueda, Yamaoka, 
\& Remillard 2009; Cui 1999; Belloni et al. 2006). 

The much longer timescale of the AGN QPO presents an opportunity to
examine this feature on the level of individual cycles. 
This contrasts with the high frequency BHB QPOs which can
currently only be studied in a statistical sense by averaging the
behaviour over very many cycles, due to the small number of photons
detected per cycle. Thus the AGN QPO may give more stringent tests of
the origin of this oscillation than the corresponding BHB data.

Both the spectrum and variability of these data can be well modelled
using two components, namely an extremely large soft X-ray excess (SX)
which remains relatively constant on the timescale of the QPO,
together with a weak power law tail to higher energies which is
strongly modulated on the QPO timescale (Middleton et al. 2009).
Models of the complete spectral energy distribution (SED) of this
object show that the SX is the dominant component of the total
bolometric luminosity of $\sim$5$\times$10$^{44}$erg s$^{-1}$
(Puchnarewicz et al. 1998; Wang \& Netzer 2003; Casebeer et al. 2006;
Middleton et al.  2007; Jin et al. in press), giving L/L$_{Edd}$ $\sim$
4-1 for the black hole mass range of 1-4$\times$ 10$^6$M$_{\odot}$
discussed above.  Thus the equivalent sources in terms of high
Eddington fraction are GRS 1915+105 (see above), which has been
accreting at or above Eddington for more than 20 years (e.g. Truss \&
Done 2006) and probably also the Ultra-luminous X-ray Sources (ULXs;
see Fabbiano 1989; Miller \& Colbert 2004; Roberts 2007).  Both of
these can show an SED dominated by a similarly distorted `hot disc'
component to that which forms the SX in RE J1034+105 (Zdziarski et
al. 2003; Middleton et al. 2009; Gladstone, Roberts \& Done 2009).

\begin{figure*}
\begin{center}
\begin{tabular}{l}
 \epsfxsize=17cm \epsfbox{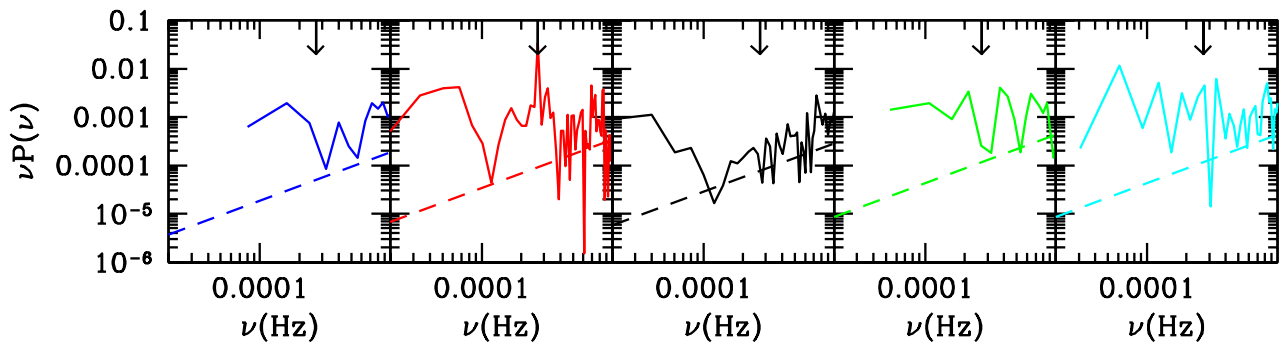}
\end{tabular}
\end{center}
\caption{The PDS of Obs1 (blue), Obs2 (red), Obs3 (black) and Obs4 (a:
green and b: cyan) made from a single realisation of the variability
(thus no error-bars are shown as they are, by definition, the size of
the power itself), including statistical white noise, the level of
which is indicated by the dashed line. The PDS of the follow-up
observations do not show the highly significant QPO seen at
$\sim$2.7$\times$10$^{-4}$Hz in Obs2, indicated by the arrows.}
\label{fig:l}
\end{figure*}

\begin{table*}
\begin{center}
\begin{minipage}{90mm} 
\bigskip
\begin{tabular}{c|c|c|c|c}
  \hline
  Date     &  OBSID &   Full exposure   &   Good exposure  & Mode \\
        & & (to nearest ks) & (all detectors) & \\
   \hline
01-05-2002 & 0109070101 & 16 & -- & FW\\
31-05-2007 & 0506440101 & 93 & 84 & FW\\
31-05-2009 & 0561580201 & 70 & 50 & SW\\
09-05-2010 & 0655310101 & 52 & 20 & SW\\
11-05-2010 & 0655310201 & 54 & 30 & SW\\

  \hline

\end{tabular}

\end{minipage}
\caption{Useful information about each of the five {\it XMM-Newton}
observations. FW indicates that the observation was taken in
Full-Window mode and SW indicates that Small-Window mode was used.}
\end{center}
\end{table*}

In this paper we analyse three new observations of RE J1034+396
obtained by {\it XMM-Newton}, together with the previous two data sets
(including the one with the QPO detection). The QPO is plainly not
detected in the first and third of the new datasets, but we may be seeing
residual signs of its presence in the second. We look at the
corresponding spectral evolution of the source, to try to identify
what triggers the appearance (and disappearance) of the QPO.

\section{Data extraction}

In this paper we will compare the spectral and timing properties of
all five observations of RE J1034+396 with \textit{XMM-Newton}. For each
observation we use 45'' regions and extract co-added MOS and PN,
background subtracted lightcurves (using {\sc sas v9.0/10.0}) to
maximise signal to noise (see details of extraction method in
Gierli{\'n}ski et al. 2008 and Middleton et al. 2009). The dates,
OBSIDs, total exposure time, useful exposure time and operational mode
are provided in Table 1.

The first two observations (hereafter Obs1 and Obs2) were carried out
in full-window mode and so are piled up in all 3 detectors due to the
extremely soft nature of the source. Obs1 is also heavily contaminated
by soft protons, with strong background flares across almost all the
observation. Any conservative background selection would give almost
no usable data. Including all the data gives the lightcurve shown in
Fig 1a with the one low point marking the position of the largest
background flare.

Obs2 is much less affected by background and we use the same good time
intervals as Gierli{\'n}ski et al. (2008).  Fig 1 shows the $\sim$50
ks segment of the $\sim$90 ks in which the QPO is most coherent and
this feature is discussed at length in Gierli{\'n}ski et al. (2008)
and Middleton et al. (2009). 

Two further observations have been taken of RE J1034+396 since the QPO
detection, the first was through director's discretionary time in AO8
(Obs3 hereafter) where the level of flux can be seen to have risen, and
the second was taken in AO9 and was broken into 2 $\sim$ 50ks
pointings (Obs4a and 4b hereafter). All of these were taken in small
window mode and so are not piled up.

\begin{figure*}
\begin{center}
\begin{tabular}{l}
 \epsfxsize=16cm \epsfbox{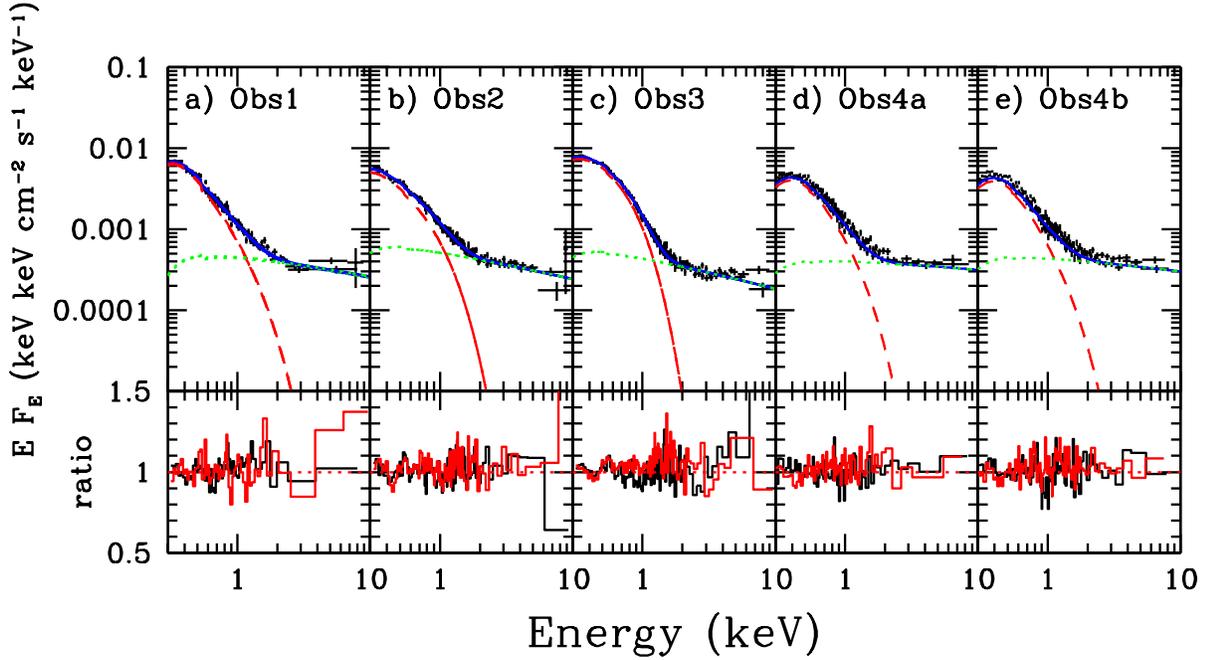}
\end{tabular}
\end{center}
\caption{Time-averaged X-ray spectra of all five observations of RE J1034+396 taken
with \textit{XMM-Newton}. The data and best-fit unfolded model are
absorbed with a minimum column equal to Galactic n$_{\rm H}$ and are
composed of a low-temperature thermal Compton component (red) and
high-temperature thermal Compton component (green) which are convolved
to give the total model (blue). Residuals to the fit are shown in the
panels below and clearly demonstrate
the presence of soft atomic features.}
\label{fig:l}
\end{figure*}

\begin{table*}
\begin{center}
\begin{minipage}{170mm} 
\bigskip
\begin{tabular}{c|c|l|l|c|c|c|c|c|c|c}
  \hline Obs & {\sc tbabs} & \multicolumn{4}{|c|}{{\sc comptt}} & \multicolumn{2}{|c|}{\sc nthcomp} &
$\chi^{2}$ (d.o.f.)\\ 
\hline 
& n$_{\rm H}\times$10$^{20}$cm$^{-2}$ & T$_0$ (keV) & kT$_e$ (keV) & $\tau$ & Flux ($\times$10$^{-12}$) &
$\Gamma$ & Flux ($\times$10$^{-12}$) &\\ 
\hline 
1 & 2.82 $^{+1.09}_{-0.85}$ & 0.038 $^{+0.015}_{-peg}$ & 0.281
$^{+0.019}_{-0.050}$ & 9.0 $^{+1.4}_{-peg}$ & 10.09 & 2.24
$^{+0.21}_{-0.24}$ & 2.33 & 227 (215)\\ 
2 & 1.47 $^{+0.12}_{-peg}$ & 0.033 $^{+0.006}_{-peg}$ &
0.195 $^{+0.018}_{-0.014}$ & 14.1 $^{+1.3}_{-1.0}$ & 6.84 & 2.33
$^{+0.08}_{-0.09}$ & 2.62 & 390 (355)\\
3 & 1.47 $^{+0.15}_{-peg}$ & 0.030 $^{+0.009}_{-peg}$ & 0.140
$^{+0.004}_{-0.003}$ & 21.5 $^{+0.9}_{-0.9}$ & 11.22 & 2.37 $^{+0.06}_{-0.07}$
& 2.19 & 511 (350)\\ 
4a & 1.47 $^{+0.64}_{-peg}$ & 0.072 $^{+0.005}_{-0.011}$ & 0.220
$^{+0.291}_{-0.063}$ & 12.3 $^{+6.5}_{-8.3}$ & 6.50 & 2.11 $^{+0.15}_{-0.47}$
& 2.18 & 223 (246) \\ 
4b & 1.47 $^{+0.003}_{-peg}$ & 0.073 $^{+0.003}_{-0.004}$ & 0.253
$^{+0.305}_{-0.081}$ & 10.1 $^{+5.3}_{-5.4}$ & 6.21 & 2.17 $^{+0.15}_{-0.13}$
& 2.31 & 340 (297)\\ 

\hline

\end{tabular}

\end{minipage}
\caption{Best-fit model parameters with 90\% confidence limits where
given. Note that the values for Obs1 are not well constrained due to
the shorter duration of the observation. Importantly however, the
higher quality data in the latter two observations gives us confidence
in our interpretation. Where the error limit is given as {\it peg},
this indicates the parameter has pegged at its hard limit. In the case
of n$_{H}$ this is the Galactic value of
1.47$\times$10$^{20}$cm$^{-2}$ , in the case of $\tau$ this is 9 (to
produce constrained fits) and in the case of T$_0$ this is
0.01~keV. The units of flux for each individual component are erg
s$^{-1}$ cm$^{-2}$ integrated over the 0.3--10 keV bandpass.}
\end{center}
\end{table*}

\subsection{Power spectra}

Figure 2 shows the corresponding PDS (plotted as
frequency$\times$power, where power is normalised to fractional
rms$^2$)  of the lightcurves shown in Figure 1.  The
100s binning of the lightcurves gives a Nyquist frequency (and hence
upper limit on the frequency on which the PDS can be determined) of
5$\times$10$^{-3}$Hz.
The PDS includes the Poisson white
noise component which starts to dominate at high frequencies, with 
normalisation given by the error bar variance on each point as shown
by the dotted line on each figure. 

While the PDS of Obs2 shows a significant QPO ($>$3$\sigma$, even when
applying the most stringent Bayesian analyses - see Vaughan 2010), the
PDS of Obs1, Obs3 and Obs 4a/b do not have a similarly significant
feature at this frequency. In particular, Obs3 sets very stringent
limits on the presence of a similarly strong QPO,
Thus the QPO is not persistent, strengthening its association with the
transient high-frequency (HF) QPOs seen in BHBs (Middleton \& Done
2010) rather than the much more ubiquitous ~LFQPOs
(Remillard \& McClintock 2006). 

We use the maximum likelihood fitting routine of Vaughan (2010),
assuming an intrinsic power law PDS ($P(f)\propto f^{\alpha}$) plus
white noise, where $\alpha$ is $-0.90\pm 0.17$, $-1.48 \pm 0.32$ and
$-1.59 \pm 0.23$ for Obs3, Obs4a and 4b respectively, compared to
$-1.80 \pm 0.20$ for Obs2 (consistent at the 2$\sigma$ level with our
more simplistic fit of -1.35 - see Gierli{\'n}ski et al. 2008).  These
indices are poorly constrained in Obs4a/b due to the relatively short
lengths of the continuous lightcurves, but Obs3 is significantly
different to Obs2. This change in red noise and QPO properties means
that the PDS is non-stationary. This is  even more 
dramatic than in NGC4051 (Miller et al. 2010), which was the
first non-stationary AGN PDS.

\begin{figure*}
\begin{center}
\begin{tabular}{l}
 \epsfxsize=16cm \epsfbox{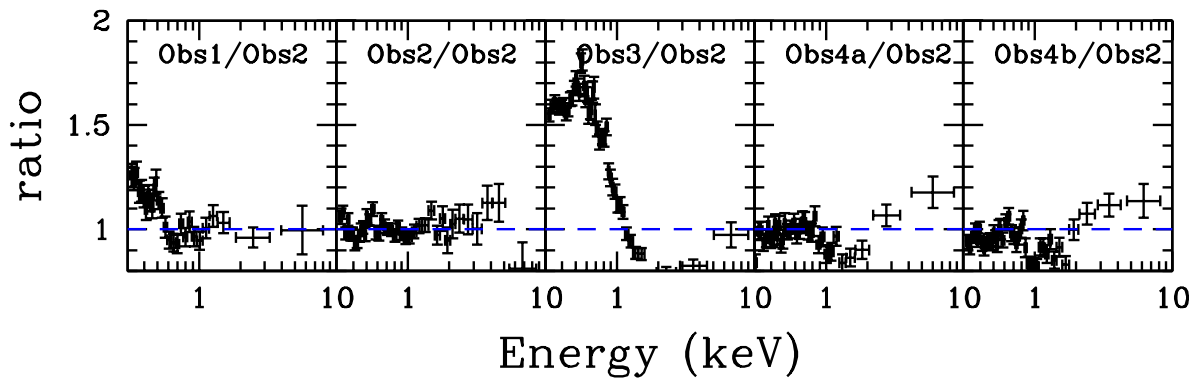}
\end{tabular}
\end{center}
\caption{Ratios of the MOS1 data to best-fitting model of Obs2. Whilst
the soft component is clearly more dominant in Obs1 and Obs3, in the 2
parts of Obs4 the soft component is consistent with that when the QPO is present.}
\label{fig:l}
\end{figure*}

\section{X-ray spectra}

As both Obs1 and 2 are heavily piled up we attempt to mitigate the
worst of the effects (i.e. migration of counts from soft to hard
energies). In the case of Obs1, centroid removal is prohibitively
expensive in terms of counts, so instead we extract the X-ray spectrum
from single patterns only. Obs2 is substantially longer and so we
remove an inner centroid of 7.5'' to account for the worst effects of
pile-up.  We extract and fit MOS spectra only, due to the mismatch
between MOS and PN spectra at low energies, using {\sc xspec} v11.3.2.

We apply the best-fitting model of Middleton et al. (2009): a
low-temperature, optically thick thermal Comptonisation of
(unobservable UV/EUV) disc seed photons, together with
high-temperature, optically thin Comptonisation which dominates at
high energies (in {\sc xspec} this is {\sc
tbabs*(comptt+nthcomp)}). The low temperature Comptonisation component ({\sc comptt} here)
is similar to that invoked for high signal-to-noise ULX (Gladstone et
al. 2009) which, if composed of high stellar mass black holes ($\le$
100M$_{\odot}$) rather than intermediate mass BHs, would share similar
Eddington/super-Eddington mass accretion rates as inferred for
RE J1034+396. We fit this model to the spectra of all five
observations (Fig 3), with model
parameters and their 90\% confidence limits presented in Table 2.

Fig 4 shows the ratio between the MOS1 data for each non-QPO
observation to the best fitting spectral model for Obs2.  These ratio
plots show that the soft component is relatively stronger in Obs1 and
Obs3 than in Obs2, but Obs4a/b have spectral shapes which are very
similar to that seen when the QPO was observed though with apparant stronger warm absorber features at $\sim$ 0.9--2~keV. The SX variability is
clearly associated with an intrinsic change rather than decreased
absorption as the column density pegs at the lower limit of absorption
in our Galaxy (1.47$\times$10$^{20}$cm$^{-2}$, taken from the n$_{\rm H}$ tool\footnotemark) except for Obs1 where
pileup/background issues mean the spectra are not reliable. 
Whilst the first three
observations appear to have broadly consistent parameters
for the shape of the components, their normalisations differ, with
Obs2 having a lower SX component (see also Fig 2). Conversely, 
Obs4a and 4b appear to have an increased seed photon and electron
temperature for the SX, with a less steep tail to high energies,
though the ratio of flux in SX and tail is similar to Obs2 (see also
Fig 2). 


\footnotetext {http://heasarc.gsfc.nasa.gov}

\begin{figure*}
\begin{center}
\begin{tabular}{l}
 \epsfxsize=10cm \epsfbox{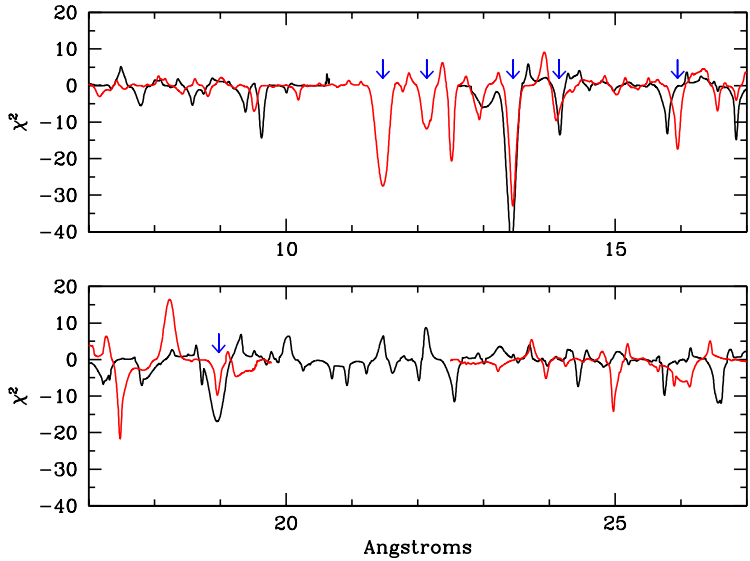}
\end{tabular}
\end{center}
\caption{Co-added RGS spectra from Obs2 and Obs3 showing each
respective arm (arm 1 in black, arm 2 in red), with the best fitting
Gaussian lines from the algorithm of Page et al. (2003), corrected for
the source redshift of 0.042. The properties of each line marked with
an arrow is given in Table 3.}
\label{fig:l}
\end{figure*}

\begin{table*}
\begin{center}
\begin{minipage}{85mm} 
\bigskip

\begin{tabular}{l|c|c|c|c|c}
  \hline

species & $\lambda_{\rm obs} ($\AA$) $ & $\lambda_{\rm lab} ($\AA$) $ &  z (km s $^{-1}$) & EW $($\AA$)$ & $\sigma$ (km s$^{-1}$)\\
   \hline
   Fe XIX   & 13.43 & 13.53 & -2200 $\pm$ 350& 0.071 & 475\\
   Fe XVIII & 14.13 & 14.21 & -1700 $\pm$ 300& 0.049 & 300\\
   Fe XVIII & 15.87 & 16.08 & -3900 $\pm$ 300& 0.078 & 225\\
   OVIII    & 18.96 & 18.97 &  -200 $\pm$ 250& 0.047 & 175\\
   Fe XXII$^{*}$  & 11.47 & 11.77 & -7600 $\pm$ 400& 0.096 & 2000\\
   Fe XXI$^{*}$   & 12.14 & 12.28 & -3400 $\pm$ 350& 0.059 & 1850\\
   \hline

\end{tabular}

\end{minipage}
\caption{Average line parameters from running the Gaussian line fitting
algorithm of Page et al. (2003) where both arms of the detector
register a significant feature ($\Delta\chi^{2} >$ 9) or it is significantly detected
in one over a gap in the other (denoted by $^{*}$). The
$\lambda_{\rm obs}$ values are corrected for the source redshift of
0.042 and are compared to the rest frame wavelengths of the ionised
species to determine the Doppler shift of the absorbing material (given to the nearest 100 km s$^{-1}$). The errors on the redshift velocities are given to the nearest 50 km s$^{-1}$ and are determined assuming the an error on the observed wavelength of 0.15 m$\AA$, i.e. half the width of the spectral binning.}
\end{center}
\end{table*}

\subsection{RGS spectra}

The flux ratios in Fig 4 and residuals to the spectral fits (lower
panels in Fig 3) indicate the presence of soft absorption features
which may be consistent with a `warm absorber' (Kaastra et al. 2000;
Kaspi et al. 2000a). We use the RGS to search for atomic features
associated with this, but the poor signal-to-noise means that we need
to co-add datasets. The previous section has already shown that there
are significant spectral changes between observations, and even in a
single observation (Obs2) there are subtle changes in the spectrum
with time. In particular, the dip in the lightcurve of Obs2 (Fig1b) at
$\sim$ 40-50~ks, possibly due to an occultation event as seen in other
AGN (McKernan \& Yaqoob 1998; Gallo et al. 2004; Risaliti et al. 2007;
Turner et al. 2008), has a clear energy dependence (Figure 1 of
Middleton et al. 2009), while Maitra \& Miller (2010) see subtle
spectral changes in atomic features correlated with the
QPO. Nonetheless, due to the limited statistics, we choose to co-add
all the data from Obs2 and Obs3 in order to get the most sensitive
limits on the time averaged spectral properties. 

We analyse the RGS spectrum in detail, fitting individual Gaussian
lines to the spectrum, following the algorithm of Page et al. (2003)
with equivalent width (EW), $\sigma$ and centroid wavelength as free
parameters. The results of this are shown in Fig. 5 for each arm of
the RGS separately (red and black), with wavelength scale corrected
for the source redshift. We show the fit parameters along with likely
identification in Table 3 for lines seen in both arms at $>$3$\sigma$
significance ($\Delta\chi^{2}>$ 9), or for lines at this significance
which are seen in only one arm if the other arm has no data at that
wavelength range. We ignore any features close to data gaps.  The
majority of these features have a small outflow velocity, of order
1000s~km/s, showing that they are most probably associated
with a wind from the broad line region/torus. Material from closer in
to the accretion disc would most probably have much higher velocity.
The line widths also imply a similar turbulent velocity of $\sim
400-1000$~km/s, again arguing strongly against an inner disc origin
for this material.

We fit these features with an ionised absorption model, {\sc xstar}
(grid25BIG with 200 kms$^{-1}$ turbulent velocity - Reeves et al. 2008),
together with an absorbed, low-temperature Compton component to
account for the continuum. The inclusion of a column of
2.23$\times$10$^{21}$cm$^{-2}$ and ionisation, log$\xi$ = 2.7 gives an
improvement in $\chi^2$ of 116 for 2 d.o.f, where the strongest
predicted features are from ionised populations of Oxygen, Neon and
Iron. This column and ionisation parameter are similar to those
derived by Maitra \& Miller (2010) from the lower resolution EPIC
spectra of Obs2, but their time resolved analysis suggested that these
absorption features changed with the QPO period, so they suggest that
the QPO is caused by an orbiting cloud. However, the $\sim 3700$~s
period coupled with the likely mass of 2$\times$10$^{6}$M$_{\odot}$
requires that the cloud is at a distance of only $15~R_g$. This is
completely incompatible with the relatively low velocity of this
material seen in both the turbulence and outflow in these higher
resolution RGS spectra.

Maitra \& Miller (2010) suggest that a change of $>$30\% in ionistion
fraction would be required to explain the changes seen based on the
phase binned EPIC spectra of Obs2. We perform a similar phase binned
selection process with the higher resolution (although lower SNR) RGS
data of Obs2 alone and fit this with narrow absorption of the soft
continuum and find that the major difference between the two data sets
is the change in continuum normalisation (at $\sim$ 6\% as expected
from the periodic behaviour) with a $<$3$\sigma$ difference in $\Delta\chi^2$ when the ionistaion parameters are the
same (log$\xi \sim$ 3 and overall fit quality of 903/898). This suggests that, rather than an absorbing column creating the
QPO, the correlated absorption features are instead due to material in
the line of sight responding to a changing illumination. This sets
only a limit on its density, that $n>7\times 10^{7}$ cm$^{-3}$, in
order that the material can recombine on timescales substantially less
than the period (e.g. McHardy et al 1995).


\begin{figure*}
\begin{center}
\begin{tabular}{l}
 \epsfxsize=17cm \epsfbox{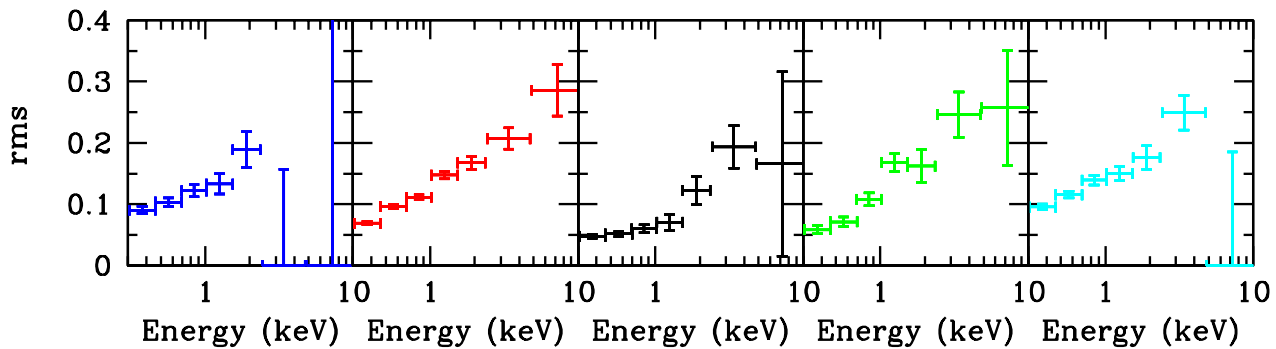}
\end{tabular}
\end{center}
\caption{rms spectra of all four observations made using co-added,
background subtracted lightcurves. Whilst pile-up will have an effect
on the observations taken in full window mode (Obs1 - blue and Obs2 -
red) this should only provide a small constant offset as the
variability transfer is from soft energies where the stable spectral
component dominates.}
\label{fig:l}
\end{figure*}

\section{Energy-dependent variability}

The PDS and mean X-ray spectra allow us to quantify the variability and
spectral behaviour when averaged over energy and time respectively.
However, it is possible to use more sophisticated spectral-timing
tools which explore the amplitude of variability as a function of
energy.

\subsection{RMS spectra}

The fractional root-mean-square variability amplitude (rms hereafter:
Edelson et al. 2002; Markowitz, Edelson \& Vaughan 2003; Vaughan et
al. 2003) gives a simple view of the energy-dependence of the
variability. In essence, it involves taking a lightcurve in each
energy band and then plotting the fractional rms of each light curve
as a function of energy. This was done for the Obs2 data in Middleton
et al. (2009), where the strong rise in rms as a function of energy
almost exactly mirrored the ratio of hard power-law to soft excess
component derived from the X-ray spectrum. This lends support to the
two-component spectral decomposition, with the two components having
different variability properties. Most of the short term variability
is associated with the power-law component, but this is increasingly
diluted at low energies by the increasing contribution of the mostly
constant soft component.

\begin{figure}
\begin{center}
\begin{tabular}{l}
 \epsfxsize=8cm \epsfbox{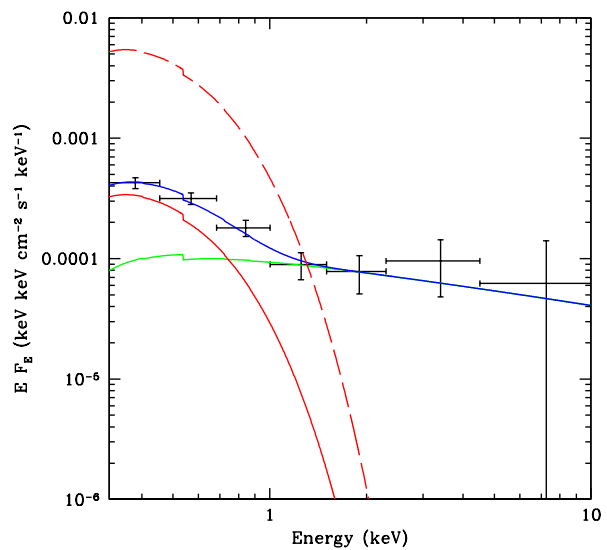}
\end{tabular}
\end{center}
\caption{Spectrum of the absolute rms of Obs3 made by multiplying the
rms by the mean counts in each band (summed over the 3 detectors) and folding with a suitably
binned spectral response. We apply the best fitting spectral model of
cool thermal Comptonisation (red) together with hot thermal
Comptonisation (green) which convolve to give the total spectrum
(blue) which is absorbed by a neutral column. The best-fit differs
from the time-averaged spectrum in having a greater relative
contribution from the hard component as can be seen by the predicted level of the soft component assuming the same contribution as the X-ray spectrum (red dashed line). The size of the error-bars demonstrates that this method of analysing the variability properties of each spectral component is very poorly
constraining due to the low signal-to-noise.}
\label{fig:l}
\end{figure}

We show the fractional rms spectra of all 5 observations in Fig 6
using 100s binning for the co-addded (MOS and PN), background
subtracted lightcurves in each energy band.  We use the total Obs2
lightcurve, not just the segment where the QPO is `in-phase', and do
not exclude the central section of the image to mitigate pileup in
either Obs1 or Obs2 as this leads to an unacceptable loss of counts.

All the rms spectra show a broadly similar shape, with rms increasing
as a function of energy, but with subtle differences in terms of the
gradient at high and low energies. To quantify this we fit a straight
line function to the log-linear data and determine the reduced
$\chi^2$ in each case. These are 0.62, 1.04, 2.47, 1.43 and 0.90 for
the 5 observations respectively and demonstrate that the greatest
deviation from the shape of the fractional variability in Obs2 is Obs3
due to the increase in the SX component.

We convert the fractional rms spectra into absolute rms spectra in
counts units by multiplying by the mean count rate (from the summed
lightcurves over all 3 detectors) in each band, and then using the
instrument response to fit this in {\sc xspec}.  This shows the
variable spectral shape directly (e.g. Revnivtsev et al. 2006;
Sobolewska \& Zycki 2006). Fig 7 shows this for Obs3, where the dashed
line shows a SX scaled so that the integrated flux is equal to that of
the time-averaged component relative to the power-law flux. Plainly,
the SX seen in the rms variability is much smaller (by a factor 16)
than this, but is consistent with the same shape.  The flux in the
variable hard component is 1.2$\times$10$^{-12}$ erg s$^{-1}$
cm$^{-2}$, compared to 2.5$\times$10$^{-12}$ erg s$^{-1}$ cm$^{-2}$ in
the variable part of the soft component. Thus while some of SX
variability could be driven by reprocessing of the illuminating hard
tail, this is unlikely to be the explanation for all of it.

\begin{figure*}
\begin{center}
 \epsfxsize=10cm \epsfbox{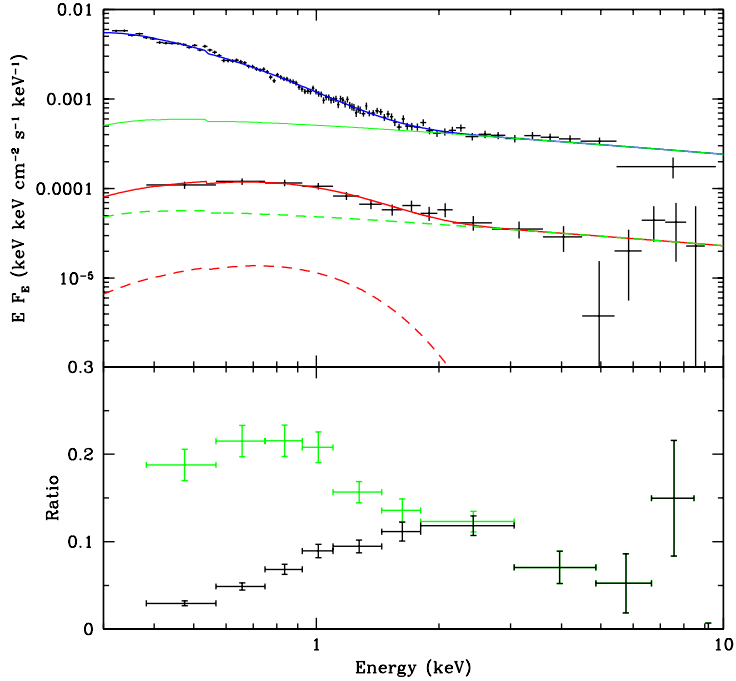}
\end{center}
\caption{{\it Upper panel}: time-averaged Obs2 spectrum showing the MOS1
data together with the best-fitting model in blue and the power-law
component in solid green. The covariance spectrum at the QPO frequency
is shown with its best fitting model in red and power-law component in
dashed green. {\it Lower panel}: The shape of the coherent rms at the QPO
frequency can be re-extracted by obtaining the ratio of the covariance
to the time-averaged spectrum (black). As seen in Middleton et
al. (2009) the shape increases up to $\sim$ 3~keV but after this is
poorly constrained. By obtaining the ratio of the covariance data to
the power-law component of its best-fitting model (green) we can see
the shape of the variable SX. This peaks at a significantly higher
temperature than the stable time-averaged SX which suggests that this
is a different component.}
\end{figure*}

\subsection{Covariance spectra}

We can improve upon the poor statistical quality of the rms spectrum
by measuring the covariance (Wilkinson \& Uttley 2009).  This method
works by cross-correlating each narrow energy band with a much broader
reference band, which effectively acts as a matched filter to pick out
correlations, so greatly reducing the noise.  By dividing the measured
covariance by the absolute rms of the reference band, we can obtain a
covariance spectrum in detector counts, similar to the absolute rms
spectrum but with much smaller error bars.  Provided that the
underlying variations between bands are spectrally coherent
(i.e. well-correlated), this technique is so effective at improving
the statistics of the rms spectrum that the variability can be sampled
over multiple frequency band-passes by selecting bin sizes and length
of segments over which the rms is averaged). We extract the covariance
spectra (relative to the 0.36-2.16 keV band) from the PN data as this
provides us with the highest quality data from an individual
detector. Guided by the results in Middleton et al. (2009), we split
Obs2 into three frequency bands, covering long timescales
(0.01-0.2~mHz), the QPO (0.2-0.4~mHz) and short timescales
(0.4-2~mHz). By comparing these covariance spectra, it becomes
possible to determine which spectral component is dominating the
variability over a given frequency range, e.g. at the frequency of the
QPO.

\begin{figure*}
\begin{center}
\begin{tabular}{lc}
 \epsfxsize=8cm \epsfbox{rej_pulse4.ps}
&
 \epsfxsize=8cm \epsfbox{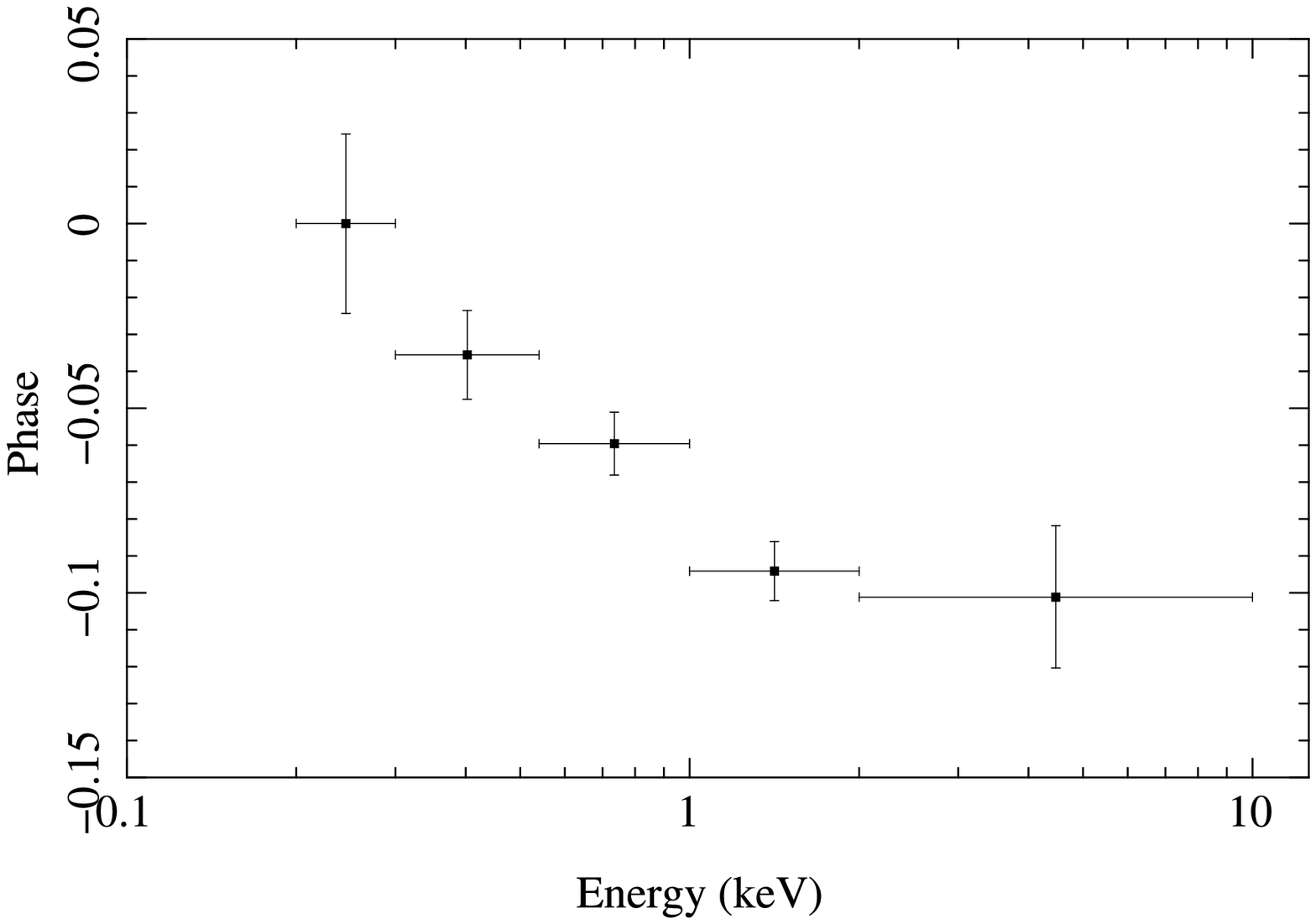}
\end{tabular}
\end{center}
\caption{{\it Left}: shows the lightcurve folded on the QPO period in different
energy bands. The dotted line suggests that the hard (1-10 keV)
emission leads the soft (0.2-0.3 keV) emission by $\sim 370$~s. {\it Right}: the
phase-energy plot of the folded lightcurves showing the constrained
phase shift of 0.1$\pm$0.01 cycles above 1~keV.}
\end{figure*}

Fig 8a shows the spectrum in the QPO band (red) compared to the time
averaged spectrum (blue). The shape is similar to that of the rms
spectrum from Obs3 shown in Fig 7 in that it is steeper than the power
law tail in the mean spectrum. A single power law fit has index is
$\Gamma=3.2\pm 0.3$ compared to $2.3$ in the mean spectrum, and also
requires significant absorption of $N_H=0.10\pm 0.05$~cm$^{-2}$ in
order to fit the turn-down at the lowest energies. While this is a
good fit, with $\chi^2_\nu=40/50$, the energy dependence of the QPO
supports instead a two component model. Although folding aperiodic
noise on a given period can produce misleading results (Benlloch et
al. 2001) the red noise is dominated by the periodic process on the
QPO frequency in Obs2. Fig 9a shows the resulting folded lightcurve of
the QPO as a function of energy, i.e. the lightcurve in a given energy
band folded on the QPO frequency. These suggest that there is a small
lag from hard to soft energies which is quantified in Fig 9b which
shows that the lightcurve above $\sim$1~keV leads the soft bands by
$\sim$0.1$\pm$0.01 cycles, hence we fit the two component model which
fits the mean spectrum. We first let only the normalisations of these
two components be free but this results in an unacceptable fit with
$\chi^2_\nu=72/51$, with residuals concentrated at low energies.  We
additionally allow the shape of the soft excess to be free and find a
much better fit ($\chi^2_\nu=38/48$) for significantly higher electron
and seed photon temperatures of $kT_e=0.31 \pm 0.10$~keV and
$kT_0=0.12\pm 0.05$~keV,respectively. Alternatively, a single
blackbody component with $kT=0.17\pm 0.03$~keV also fits the shape of
the soft component in the QPO ($\chi^2_\nu=39/49$).

Thus one attractive possibility is that the soft lag is due to
reprocessing, with the QPO in the hard tail also giving a QPO in the
illuminating spectrum seen by the accretion disc. The energetics are
again an issue as the reflection albedo of the disc should be quite
high, however, with the soft component flux of 2.6$\times$10$^{-13}$
ergs cm$^{-2}$ s$^{-1}$, comparable to 4.3$\times$10$^{-13}$ ergs
cm$^{-2}$ s$^{-1}$ of the hard component it appears plausible that the
soft component is made via reprocessing.

Fig 10 shows the covariance spectra at the QPO timescale (0.2-0.4~mHz:
black) compared to that of the longer (0.01-0.2~mHz: red) and shorter
(0.4-2~mHz: green) timescale variability. The rapid variability is
somewhat harder than the QPO but still contains a similar but smaller
hot, soft component. This subtle difference between the QPO and rapid
variability is not at all evident from the rms spectra alone
(Middleton et al 2009). It is only revealed by these better
signal-to-noise covariance spectra. However, the change in spectral
shape between the QPO and long timescale variability is so large that
this is clear even in the rms spectra (Middleton et al 2009).

The remaining observations have insufficient variability power at the
QPO frequency to make covariance spectra in the narrow QPO energy
band, so we use only two frequency bands, covering long (0.02-0.2~mHz)
and short (0.2-2~mHz) timescale variability, and compare these to the
same frequency bands in Obs2, where 'short' corresponds to the
co-added QPO and rapid variability spectrum.  The short timescale
covariance spectra are shown in Fig 11a. Obs4a and b are very similar
to Obs2, showing the hotter soft excess component which is present in
the QPO however Obs3 (black) is slightly softer, appearing to peak at
lower temperatures consistent with the SX in the mean spectrum.  Fig
11b shows the corresponding plot for the long timescale variability,
showing clearly that this is dominated by the soft excess in all
observations with a shape similar to the SX component in the
time-averaged spectrum.

\section{Discussion}

Understanding the variability of RE J1034+396 is important as the
significant QPO (in Obs2) makes it unique amongst AGN. The QPO is
plainly not present in Obs3, showing that the QPO is transient rather
than a long-lived feature of the lightcurve. Comparing the spectral
and timing properties of the average and frequency binned data allows
us to compare and contrast the physics whilst the QPO was present and
not present. This also allows us to test some of the proposed models
for the QPO (e.g. Maitra \& Miller 2010; Das \& Czerny 2010).

All of the observations are consistent with a two component continuum,
with a strong soft excess and weak high energy tail. The strength and
shape of the soft excess clearly varies between observations, while
the (much lower signal to noise) power law tail remains fairly
constant. However, on much shorter timescales (within a single
observation) this behaviour is reversed.  The fractional r.m.s as a
function of energy shows that the soft excess is always much less
variable than the tail, especially on the most rapid timescales. The
dominance of the soft excess in the energy spectrum means that this
strongly dilutes the amount of rapid variability at low energies as in
Middleton et al (2009).

However, we now go beyond this and look in detail at the spectrum
of the variability using the much better signal-to-noise covariance 
spectra. This reveals a new aspect of the QPO itself 
(0.2-0.4~mHz variability i.e. 5000-2500s timescale), which is that, as 
well as the power law tail, it also has additional variability at 
lower energies. This component is clearly hotter than the 
soft excess seen in the mean spectrum, and the reality of this 
as an additional component is supported by it lagging 
behind the 2-10~keV variability by ~370~s. This lag could be produced by
reprocessing, in which case it corresponds 
to a light crossing distance of $30R_g$ for a 
$2\times 10^6 M_\odot$ black hole.

This new information on the QPO spectral shape puts some constraints
on models. For example, a radial perturbation of the Compton torus
leads to an offset geometry i.e. to elliptical orbits whose perihelion
precesses. However, there is no change in illumination of the disc
during the azimuthal precession so it is hard to incorporate a soft
lag from reprocessing into this model. Conversely, a vertical
oscillation or coupled vertical-radial oscillation clearly changes the
disc illumination (Abramowicz \& Kluzniak 2005; Blaes et al 2007).

\begin{figure}
\begin{center}
\begin{tabular}{l}
 \epsfxsize=8cm \epsfbox{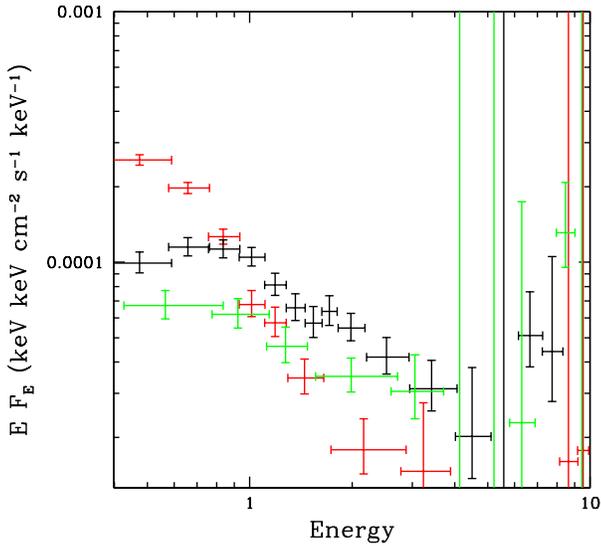}
\end{tabular}
\end{center}
\caption{The covariance spectra from Obs2 on short (green: 0.4-2~mHz),
QPO (black: 0.2-0.4~mHz) and long (red: 0.01-0.2~mHz) timescales. The
short timescale variability is similar to that of the QPO, but has
less of the hot soft component in its spectrum. The long timescale
variability is similar in shape to the soft excess seen in the mean
spectrum.}
\label{fig:l}
\end{figure}

An alternative set of models incorporate a transonic shock rather than
a global mode. The shock could occur in a hot, very low angular
momentum flow, and its radial position can oscillate (Das \& Czerny
2010). The maximum in hard X-ray flux is where this shock radius is
largest, which leads also to stronger illumination of the disc, so
these models can also produce a soft lag from reprocessing.

The QPO spectrum is also clearly not affected by strong absorption
features. An absorption origin for the QPO is also ruled out by the
RGS spectrum which shows that the atomic features (which are present
in the spectrum as claimed in Maitra \& Miller 2010) are narrow and
not strongly Doppler shifted. The small distances required for
periodic obscuration by an orbiting clump of material to produce the
QPO would imply much larger velocities. The periodic change in
absorption seen by Maitra \& Miller (2010) may instead simply be the
wind at large distances changing in ionisation in response to the
changing illuminating flux.

None of the remaining observations show any significant excess of
power over the red noise at the QPO frequency. The one with the most
power at this frequency is Obs4b, which, along with Obs4a, has a
fairly similar short timescale covariance spectrum to that of the QPO
which may indicate a similar variablity process was occuring although
not as dominant as in Obs2.

\begin{figure*}
\begin{center}
\begin{tabular}{lc}
 \epsfxsize=8cm \epsfbox{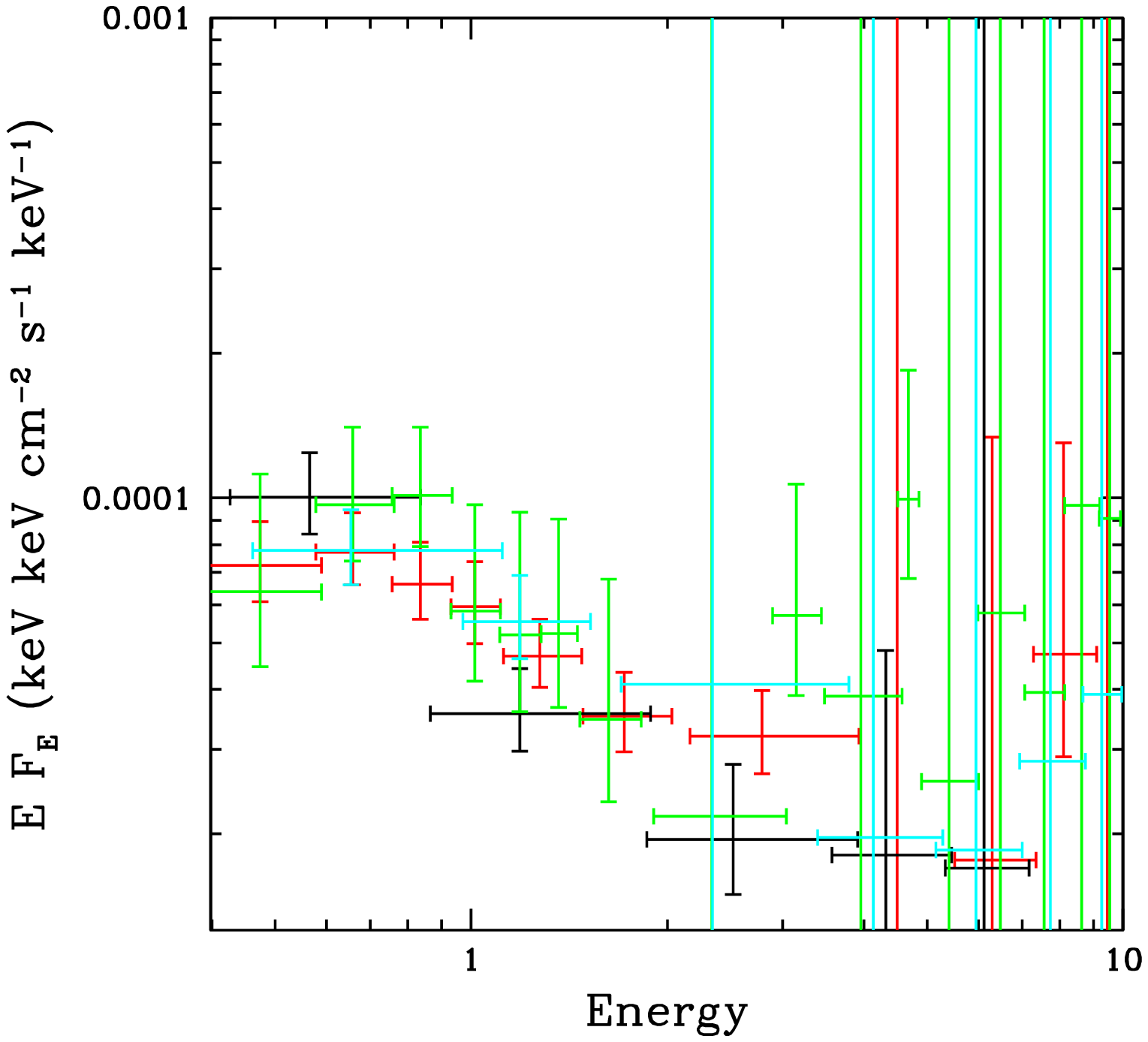}
&
 \epsfxsize=8cm \epsfbox{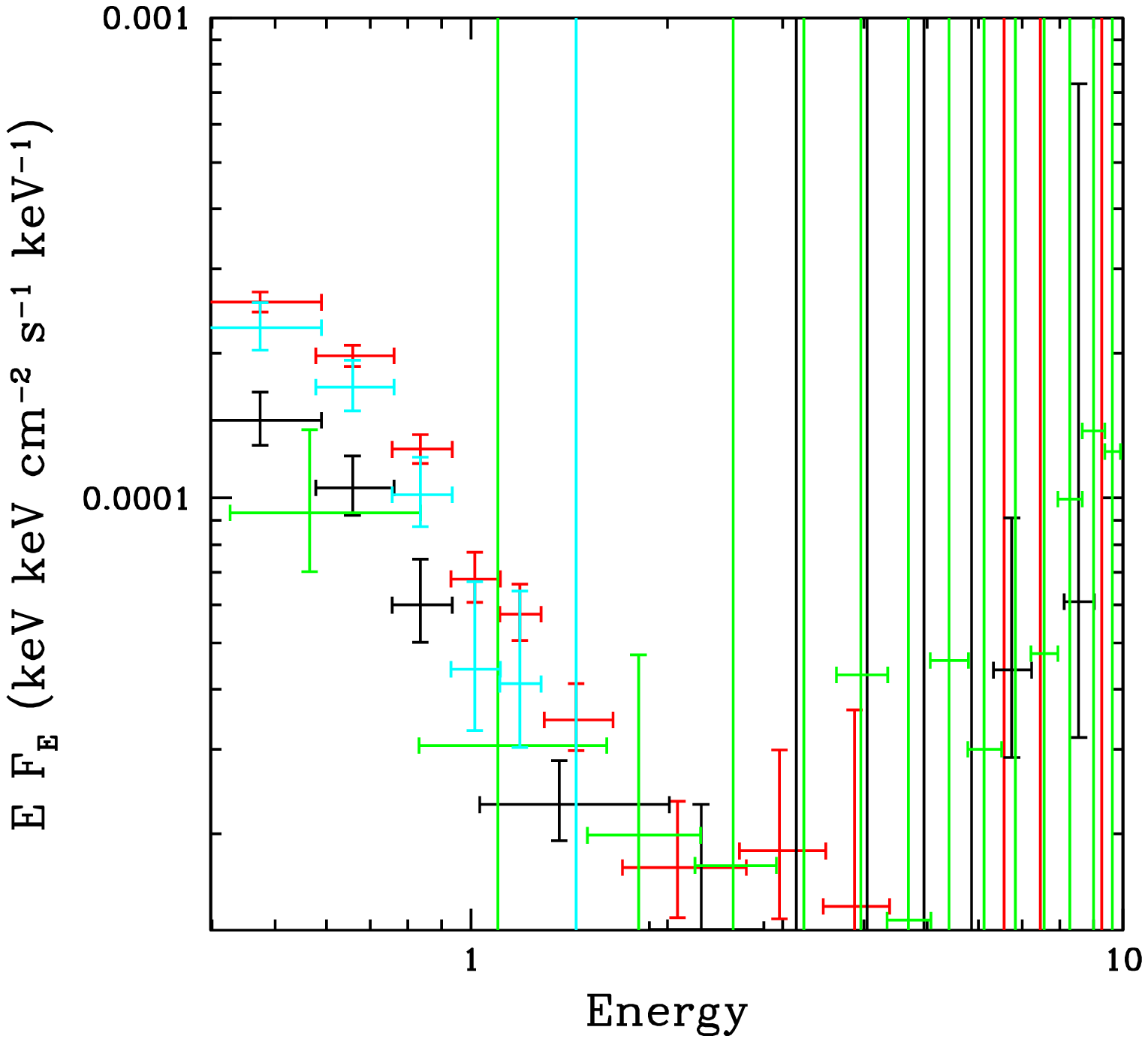}
\end{tabular}
\end{center}
\caption{{\it Left}: The covariance spectra of the short timescale
variability of the 4 well constrained observations of RE J1034+396
(red: Obs2, black: Obs3, green: Obs4a, cyan: Obs4b). The shape of the
SX variability looks like that of the time-averaged SX in Obs3
but that of Obs4 (a and to a lesser extent b) looks similar to that of the QPO
variability where a dissimilar component is present. {\it Right}: The
analogous plot of the long timescale variability where all shapes look
similar to the time-averaged SX component.}
\label{fig:l}
\end{figure*}

Thus the QPO is transient, and there is no clear trigger for its
emergence. One potential way forward is instead to assume that 
this is similar to the 67Hz QPO in GRS 1915+105 (e.g. Morgan, Remillard \& Greiner 1997) and use the
substantially larger dataset on GRS 1915+105 to search for the 
trigger of the QPO in the stellar mass black hole. 
This has not yet been done, but may prove to be illuminating. 

\section{Conclusion}

RE J1034+396 is one of the most extreme and important AGN detected
thus far as its timing properties resemble those of the highest
luminosity BHBs. Follow-up observations show that the PDS no longer
contains the strong QPO that characterised the lightcurve in the 2008
detection. Such a change in the intrinsic variability properties
indicates that the PDS is non-stationary.

We use covariance spectra to get better signal to noise on the
spectrum of the variability. This reveals new features in the QPO
itself. There is a soft component in addition to the power law tail
which contributes to the QPO signal. Folding the lightcurve on the QPO
timescale over different energy bands indicates that this component
lags behind the hard emission. This implies a size scale of $\sim
30R_g$ if this is produced by reprocessing of the hard QPO spectrum,
though we caution that the lag could also be caused by a more complex
interaction between the disc and corona, rather than by
reprocessing. However, a reprocessing origin puts some constraints on
the QPO mechanism, since this requires a changing illumination of the
disc. This probably rules out models where the QPO is from a radial
perturbation precessing around the black hole, as here there is no
change in disc illumination. Instead this could be produced by a
vertical or coupled vertical-radial (breathing mode) mode (Abramowicz
\& Kluzniak 2005; Blaes et al 2007), or the oscillating shock model of
Das \& Czerny (2010).

We rule out occultation models from an orbiting cloud from the RGS
data, which show that the ionised absorbed has narrow features at low
velocity, unlike those expected from material within $20R_g$ as
required by the QPO timescale.

None of the other 4 datasets show the QPO despite 3 of these showing
rather similar spectra and 2 of these showing a rather similar
covariance spectrum on short timescales. Thus there is no clear
trigger we can identify for the QPO from these data. We suggest
instead that it may be useful to search the much more extensive
datasets in black hole binaries for the trigger of their transient
high frequency QPOs.

\section{Acknowledgements}

We thank the anonymous referee for their useful insights and
suggestions. MM and CD acknowledge support from STFC via a standard
grant and PU via an STFC advanced fellowship. This work is based on
observations obtained with {\it XMM-Newton}, an ESA science mission
with instruments and contributions directly funded by ESA Member
States and NASA. We thank Dr Mat Page for his assistance with the
line-finding algorithm that he generously provided and Dr Simon
Vaughan for his excellent Bayesian R routines.

\label{lastpage}

\end{document}